\documentclass[aps,prl,amsmath,amssymb,nofootinbib,reprint,longbibliography,linenumbrs,superscriptaddress]{revtex4-1}
\pdfoutput=1
\usepackage{amsmath}
\usepackage{feynmp-auto}
\usepackage{slashed}
\usepackage{siunitx}
\usepackage{graphicx}
\usepackage{multirow}
\usepackage{amssymb}
\usepackage{mathtools}
\usepackage{xcolor}
\usepackage{hyperref}
\usepackage{mathtools}
\usepackage{subfigure}
\usepackage[normalem]{ulem}

\catcode`\$=\active
\gdef$#1${\texorpdfstring{\(#1\)}{\detokenize{#1}}}

\newcommand{\ud}{\mathrm{d}}

\newcommand{\inmath}[1]{\relax\ifmmode#1\else$#1$\fi}
\newcommand{\MSbar}{\inmath{\rm \overline{MS}}}

\begin{document}

\author{Fabio Maltoni}
\email{fabio.maltoni@uclouvain.be, fabio.maltoni@unibo.it}
\affiliation
{
    Centre for Cosmology, Particle Physics and Phenomenology (CP3),
    Universit\'{e} catholique de Louvain,
    1348 Louvain-la-Neuve,
    Belgium
}
\affiliation
{Dipartimento di Fisica e Astronomia, Universit\`a di Bologna and INFN, Sezione di Bologna, via Irnerio 46, 40126 Bologna, Italy}
\author{Manoj K. Mandal}
\email{manojkumar.mandal@pd.infn.it}
\affiliation
{Dipartimento di Fisica e Astronomia, Universit\`a di Padova and INFN, Sezione di Padova,
Via Marzolo 8, 35131 Padova, Italy
}
\author{Xiaoran Zhao}
\email{xiaoran.zhao@uclouvain.be}
\affiliation
{
    Centre for Cosmology, Particle Physics and Phenomenology (CP3),
    Universit\'{e} catholique de Louvain,
    1348 Louvain-la-Neuve,
    Belgium
}

\title{Top-quark effects in diphoton production through gluon fusion at NLO in QCD}

\begin{abstract}
At hadron colliders, the leading production mechanism for  a pair of photons is from quark-anti-quark annihilation at the tree level. However, due to large gluon-gluon luminosity, the loop-induced process $gg\to \gamma \gamma$ provides a substantial contribution. In particular,  the amplitudes mediated by the top quark become important at the $t \bar t$ threshold and above. In this letter we present the first complete computation of the next-to-leading order (NLO) corrections (up to $\alpha_S^3$) to this process, including contributions from the top quark. These entail two-loop  diagrams with massive propagators whose analytic expressions are unknown and have been evaluated numerically. We find that the NLO corrections to the top-quark induced terms are very large at low diphoton invariant mass $m(\gamma \gamma)$ and close to the $t \bar t$ threshold. The full result including five massless quarks and top quark contributions at NLO displays a much more pronounced change of slope in the $m(\gamma \gamma)$ distribution at $t \bar t$ threshold  than at LO and an enhancement at high invariant mass with respect to the massless calculation. 
\end{abstract}

\maketitle

{\bf Introduction} -- The production of a pair of photons (diphoton) is one of the most important processes at hadron colliders. Not only because the final state signature is experimentally very clean, but also because of the great phenomenological relevance for Standard Model (SM) physics and beyond. 
Its differential cross section has been precisely measured at the Tevatron~\cite{
Aaltonen:2012jd,
Abazov:2013pua} 
and the LHC~\cite{
Chatrchyan:2011qt,
Chatrchyan:2014fsa}. 
The signature has provided one of the two golden channels (the other being $H\to 4 \ell$) for the discovery of the Higgs boson~\cite{Aad:2012tfa,Chatrchyan:2012xdj}. Currently, the $H\to \gamma \gamma$ decay remains one of the cleanest final states to study the properties of the Higgs boson and its production mechanisms. Being so experimentally neat, the diphoton spectrum is also scrutinized in the search of new physics at the LHC, see {\it e.g.}~\cite{
Khachatryan:2016yec,
Aaboud:2017yyg}
,
such as  peak/dip structures coming from new scalar or spin-2 resonances decays and the interference with the standard model background or more exotic features, such as multiple resonances as predicted by extra-dimensional~\cite{Antoniadis:2011qw,Baryakhtar:2012wj,Cox:2012ee} or clockwork models~\cite{Giudice:2017fmj}.

\begin{figure*}[t]
	\centering
	\subfigure[\label{fig:born}]{
		\includegraphics[width=0.14\textwidth]{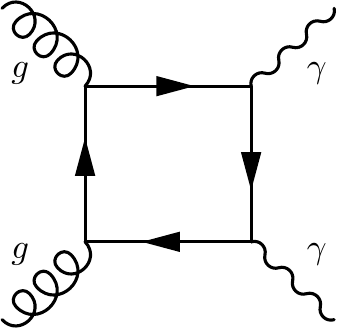}
	}
	\subfigure[\label{fig:real1}]{
		\includegraphics[width=0.14\textwidth]{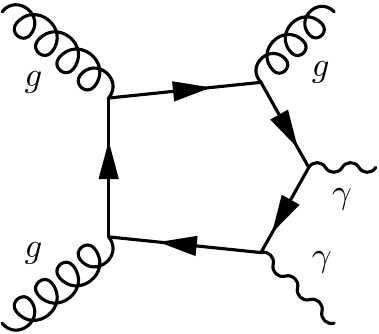}
	}
	\subfigure[\label{fig:real2}]{
		\includegraphics[width=0.14\textwidth]{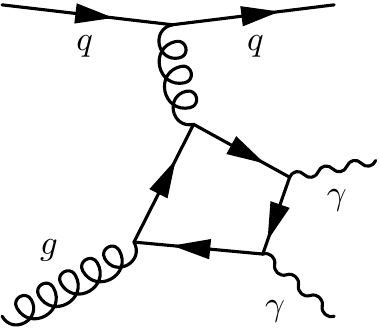}
	}
	\subfigure[\label{fig:virt1}]{
		\includegraphics[width=0.14\textwidth]{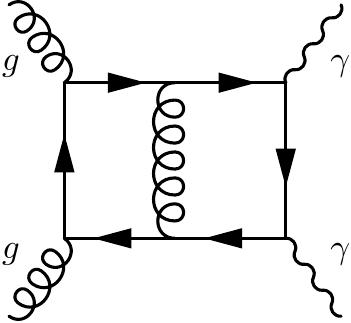}
	}
	\subfigure[\label{fig:virt2}]{
		\includegraphics[width=0.14\textwidth]{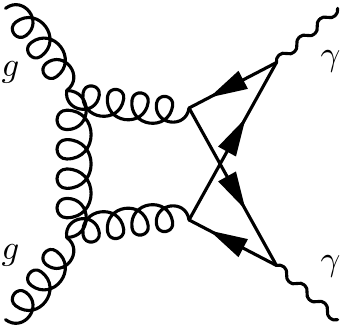}
	}
	\caption{Representative Feynman for $g g \to \gamma \gamma$ at NLO:  the Born (a), real corrections (b,c), and virtual corrections (d,e). In our computation photons couple only to the quarks running in the closed loop, {\it i.e.}, 5 massless quarks $u,d,c,s,b$ and the top quark $t$.  \label{fig:diags}}
\end{figure*}

At hadron colliders, the Leading Order (LO) contribution to diphoton final states,
comes from quark-antiquark annihilation $q\bar{q}\to \gamma\gamma$.
Next-to-Leading Order (NLO) corrections (at order $\alpha_S$) to this process have been calculated many years ago~\cite{Binoth:1999qq}. Next-to-Next-to-Leading Order (NNLO) corrections (at order $\alpha_S^2$) have also obtained~\cite{Catani:2011qz,Campbell:2016yrh} and are available in public codes such as $2\gamma$\textsc{NNLO}~\cite{Catani:2011qz}, {\sc MCFM}~\cite{Campbell:2016yrh} and {\sc Matrix}~\cite{Grazzini:2017mhc}. 
At this order, a new channel arises, {\it i.e.}, gluons can fuse into diphoton, a quantum process induced by loops of quarks (Fig.~\ref{fig:born}). This contribution, while being formally part of the NNLO corrections, is not only finite and gauge-invariant per se but also anomalously large, due to the gluon-gluon luminosity. It is common, therefore, to consider loop-induced gluon fusion production which starts at order $\alpha_S^2$ as an independent diphoton production mechanism. 
NLO corrections of this process ($\alpha_S^3$), which include two-loop $gg \to \gamma 
\gamma$ contributions,  were calculated some time ago  but only in the case of massless internal quarks~\cite{Bern:2002jx,Campbell:2016yrh}.

The top-quark contribution has been known only at one loop so far. In the low energy region, it is strongly suppressed due to the large top-quark mass, the amplitude scaling as $s^2/m_t^4$. Once the energy becomes comparable to the top mass and in particular close and above the top pair threshold, it becomes enhanced due to the opening of an imaginary part due to rescattering. This transition region is particularly interesting, because it is very sensitive to the top mass and could provide a handle on a top quark mass that is free from the usual hadronic systematic uncertainties. In the ultra-high energy limit the top quark contribution can be estimated by treating it as a massless quark: naively summing over the electric charges,
the inclusion of the top quark increases the gluon fusion contribution
by $(\sum_{6F}e_q^2)^2/(\sum_{5F}e_q^2)^2-1\approx 86\%$.

The diphoton spectrum also provides a privileged observatory to search for new physics.
While the resonant production of new physics particles decaying into diphoton can be searched with theory-independent side-band method, the non-resonant or interference cases, require a precise prediction of the SM contribution. Interference of the resonant contribution with the SM continuum, provides an important method to extract properties of the resonance, such as the width. The case of Higgs boson has been extensively investigated, see {\it e.g.}, ~\cite{Dixon:2013haa,Campbell:2017rke}. When new physics resonances are produced mostly via gluon fusion, such as for example scalars  and spin-2 particles, the SM contribution can interfere determining non-trivial structures like peak-dip (or dip-peak) or just dip structures~\cite{Chway:2015lzg}, depending on the couplings and properties of the resonance. As these new physics searches are particularly motivated above the top pair threshold, including the top-quark contribution is essential.

The computation of NLO corrections of the top-quark induced contributions, requires the knowledge of highly non-trivial two-loop amplitudes. While in the massless quark limit the corresponding amplitudes have been known for a long time~\cite{Bern:2001df}, the computation of massive ones, is still a challenge. Analytical results have become available in closed form for some of the relevant Feynman integrals (planar) yet the full set is unknown. On the other hand, numerical methods have been introduced~\cite{1707885} that allow to perform this calculation. In this letter, we compute the complete NLO corrections to the gluon fusion channel $gg\to\gamma\gamma$, including the top-quark contribution for the first time.

{\bf Calculation} -- The cross section at NLO accuracy can be written as  
\begin{align}
\ud \sigma^{\text{NLO}}=\ud \sigma^{\rm Born} + \ud \sigma^{\rm V}+\ud\sigma^{\rm R}+\ud\sigma^{\rm C}\,,
\nonumber
\end{align}
where $\ud \sigma^{\rm Born}$ is the leading order one-loop contribution, 
$\ud \sigma^{V}$ denotes the virtual (two-loop) contributions,
$\ud \sigma^{R}$ is the real (one-loop, 2$\to$3 contribution),
and $\ud\sigma^{C}$ represents the collinear singularity to absorbed into the parton distribution functions.
The representative Feynman diagrams for the Born, virtual and real contributions are shown in Fig.~\ref{fig:diags}. Each of the three terms at NLO are infrared/collinear divergent. Their sum, however,  is free of infrared/collinear divergences. To handle this cancellation, we employ an in-house implementation of the dipole subtraction method~\cite{Catani:1996vz}, which introduces counterterms for each term $\ud \sigma^{i}_{\rm fin}=\ud \sigma^{i}-\ud \sigma^{i}_{\text{dipole}}$ with $i={\rm V,R,C} \,.$
The subtraction terms $\ud\sigma^{i}_{\text{dipole}}$ are carefully chosen
such that they cancel locally the infrared/collinear divergences of each term, and sum up to zero~\cite{Catani:1996vz}.

Once the subtraction method is in place,  one is left with the calculation of the matrix elements for the virtual and real contributions.  The latter corrections require the computation of one-loop five point amplitudes, which can be done automatically. In particular, the matrix element for $g g\to \gamma\gamma g$ subprocess, as well as $g q(\bar{q})\to \gamma\gamma q(\bar{q})$ and $q\bar{q}\to\gamma\gamma g$ subprocesses are needed. To this aim, we adopt {\sc Recola2}~\cite{Denner:2017wsf} and {\sc Madgraph5\_aMC@NLO}\cite{Alwall:2014hca}, as well as analytical expression for the light-quark contributions~\cite{Bern:1993mq,deFlorian:1999tp,Balazs:1999yf}. 

We have implemented the light quark contribution from ref.~\cite{Bern:2001df} in our code. 
We have then considered the calculation of  the top-quark contribution.
Two-loop diagrams have been generated by {\sc QGRAF}~\cite{Nogueira:1991ex},
and processed by {\sc Form}~\cite{Vermaseren:2000nd,Kuipers:2012rf},
to generate corresponding amplitudes. They are fed into Reduze~\cite{vonManteuffel:2012np} to
perform the corresponding loop momentum redefinition and
to classify them into 33 integral families according to the propagator structure.
We then adopt a projection method to decompose the amplitudes into 10 independent tensor structures, reducing the computation into that of scalar integrals with irreducible numerators.
Employing the {\sc C++} version of {\sc FIRE5}~\cite{Smirnov:2014hma} with {\sc LiteRed}~\cite{Lee:2012cn, *Lee:2013mka} to perform the integration-by-part reduction, we finally obtain the corresponding form factors as a linear combination of 1180 master integrals, distributed into the 33 integral families.
We evaluate the master integrals family by family, not considering the relations among the  master integrals of different families. The calculation of the master integrals is based on numerical integration of differential equations,
with initial condition provided by an in-house implementation of sector decomposition method~\cite{Binoth:2000ps}.The numerical integration of differential equation is done with {\sc Odeint}~\cite{odeint}. 
Starting from the original initial conditions, several points in the physical region are pre-computed and results are stored. During the phase space integration, the closest point in the pre-computed set is adopted as the new initial condition. The average time to evaluate the amplitude is around $1$ second, with at least $\mathcal{O}(10^{-9})$ precision at the master integral level.
The one-loop amplitude up to $\mathcal{O}(\epsilon^2)$ order is computed within the same method.
We refer the reader to ref. ~\cite{1707885} for the details of our method and its extensive validation. Here we stress that, whenever available, we have compared the numerical value of the master integrals with those in the literature~\cite{Caron-Huot:2014lda,vonManteuffel:2017hms,Becchetti:2017abb}, and found excellent agreement.

We renormalize $\alpha_s$ in the \MSbar\ scheme with five flavors. 
The top-quark mass is renormalized on shell. 
We have checked that UV divergences are cancelled by the corresponding counter terms, and IR and collinear divergences cancel with the dipole subtraction terms. We have also checked that our implementation for the massless contribution at NLO agrees with that of MCFM~\cite{Campbell:2016yrh}\cite{note1}
%\footnote{A comparison with our results has allowed to identify two bugs in the MCFM v8.0 implementation of minor numerical relevance, which have been corrected in MCFM v8.2. In addition the change of scheme to ``tH-V'' needs to be applied to obtain full agreement with our results. This will be included in the forthcoming release of MCFM.} 
To monitor the numerical accuracy during the evaluation, we exploit the $t\leftrightarrow u$ symmetry and calculate two independent yet equivalent values of the integral in each point of the phase space.  With the uncertainties estimated through adopting different initial conditions, as well as exchanging $t$ and $u$, we conclude that the uncertainty arising from the numerical evaluation is smaller than 0.4\% times LO contribution, at both inclusive and differential level.

\begin{figure}
	\centering
 	\includegraphics[width=0.48\textwidth]{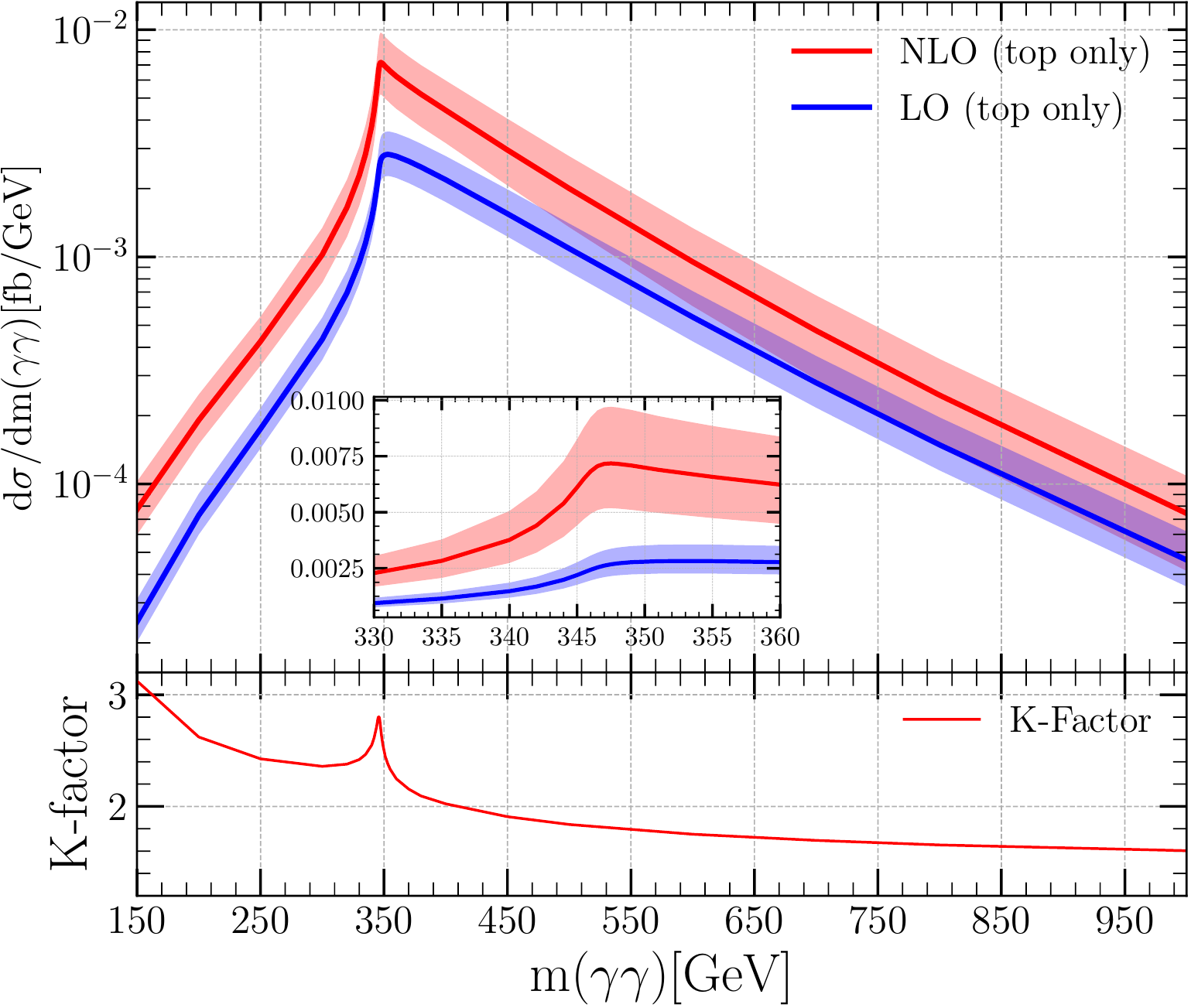}
	\caption{The differential cross section for the top only case is shown, with error bands indicate scale uncertainties.}
	\label{fig:dif-top}
\end{figure}

{\bf Results} -- We adopt the following input parameters $\alpha(0)=1/137.035999139,m_t=173.0 ~\mathrm{GeV},\Gamma_t=1.41~\mathrm{GeV}$\cite{Tanabashi:2018oca}.
We adopt the complex mass scheme for the top quark,
{\it e.g.}, the top quark mass is replaced by the complex quantity $\mu_t=\sqrt{m_t^2-i m_t\Gamma_t}$ everywhere.
The renormalization scale and factorization scale are set to $\mu_{R}=\mu_{F}=m(\gamma\gamma)/2$, and we vary them by a factor of two around the central scale to assess scale uncertainties.
We choose PDF4LHC15~\cite{Butterworth:2015oua}, and the results are presented for 13 TeV LHC. We apply the following cuts $p_T(\gamma_1)>40~ \mathrm{GeV},p_T(\gamma_2)>25 ~\mathrm{GeV},|\eta(\gamma)|<2.5$. No photon isolation is applied. 

\begin{table*}
	\caption{The differential cross section for various contribution at LO and NLO are shown for different values of diphoton invariant mass $m(\gamma\gamma)$.
		``5F only" means only including the five massless quarks,
		``top only" means only including the top quark,
		``interference" means only the interference term between the light quarks and the top quark,
		and ``full" means all the above contributions.
}
		\label{tab:dif-xs}

	\begin{tabular}{c|c|c|c|c|c|c|c|c}
		\hline
fb/GeV		& \multicolumn{4}{c|}{LO} & \multicolumn{4}{c}{NLO} \\
		\hline
		$m(\gamma\gamma)$[GeV] & full & 5F only & top only & interference & full & 5F only & top only & interference \\
		\hline
		125 & 24.26(1) & 100.1\% & $<0.01\%$ & -0.1\% & 37.3(1) & 100.1\% & $<0.01\%$ & -0.1\% \\
		400 & 0.11342(5) & 104.6\% & 1.9\% & -6.5\% & 0.1628(5) & 99.3\% & 2.7\% & -2.0\% \\
		500 & 0.03951(7) & 88.7\% & 2.8\% & 8.6\% & 0.0582(2) & 82.7\% & 3.5\% & 13.9\% \\
		1000 & \num{8.721(8)e-4} & 63.2\% & 5.3\% & 31.5\% & \num{1.266(2)e-3} & 60.5\% & 5.8\% & 33.6\% \\
		\hline
	\end{tabular}
\end{table*}
In Fig.~\ref{fig:dif-top}, we show the differential cross section for the case where only the top-quark contribution is taken into account, at LO and NLO, as well as corresponding scale uncertainties.
Both LO and NLO cross sections peak around the top-quark pair threshold.
The NLO corrections lead to a large $K$-factor~($K^{\text{NLO}}=\sigma^{\textrm{NLO}}/\sigma^{\text{LO}}$), especially in the low invariant mass region.
Even when the invariant mass $m(\gamma\gamma)$ is low,
the total center of mass energy in the real correction can be above the top pair threshold,
and thus the top-quark loop can get resolved, leading to such enhancement.
Furthermore, the photon $p_T$ cuts enhance the real corrections
since at LO photons are back-to-back and therefore have both $p_T>40\, \mathrm{GeV}$,
while at NLO the second photon can be softer.
As $m(\gamma\gamma)$ increases, the $K$-factor decreases, reaching a local maximum of value around $2.8$ at the top-quark pair threshold. In such a region, the top quarks in the loop are produced on shell and almost at rest, and can exchange a Coulomb gluon  leading to an enhancement (tamed by the top-quark width). Such corrections are well known and universal. They can be resummed by employing bound state techniques, see {\it e.g.}~\cite{Kawabata:2016aya}, though we present only fixed-order results here.

\begin{figure}
	\centering
	\includegraphics[width=0.48\textwidth]{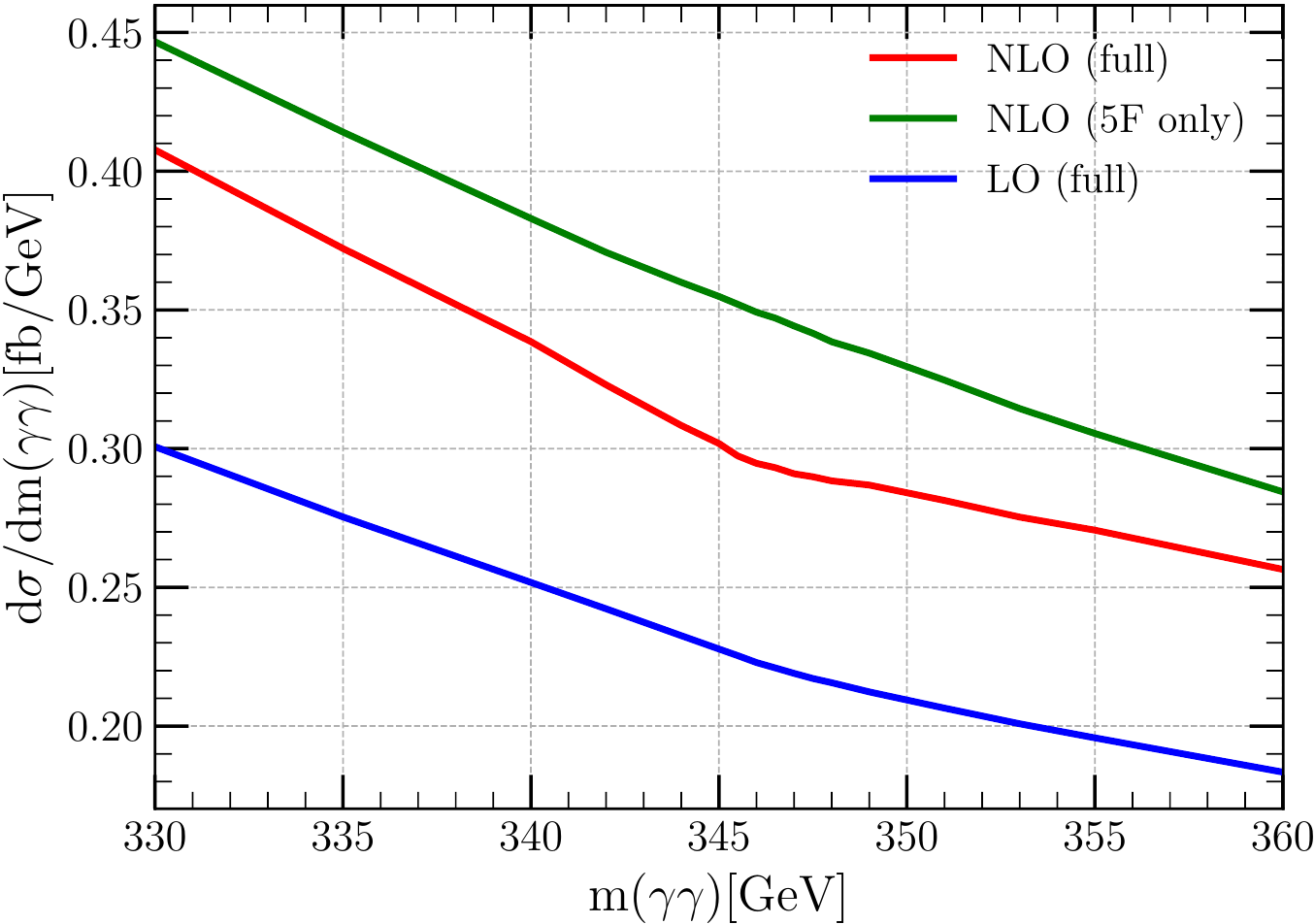}
	\caption{The differential cross section before and after including the top-quark contribution is shown for the top pair threshold region.}\label{fig:dif-top-thres}
\end{figure}

For a better view of the top-quark contribution, in Fig.~\ref{fig:dif-top-thres}
we show the differential cross section close to the top pair threshold region.
Here, the top-quark and the five massless quarks contributions
have a different phase, leading to a destructive interference that decreases the cross section.
As already mentioned, the exchange of a Coulomb gluon leads
to large NLO corrections at the top-quark  pair threshold.
Thus, the destructive interference decreases the cross section further and the change of slope below and above twice the top mass is more visible at NLO than at LO.
Our results provide a key ingredient for improving the resummed predictions in the threshold region and also  
reinforce the hope that such slope change could be exploited to extract a short-distance (potential) mass for the top quark~\cite{Jain:2016kai}.

\begin{figure}
	\centering
	\includegraphics[width=0.48\textwidth]{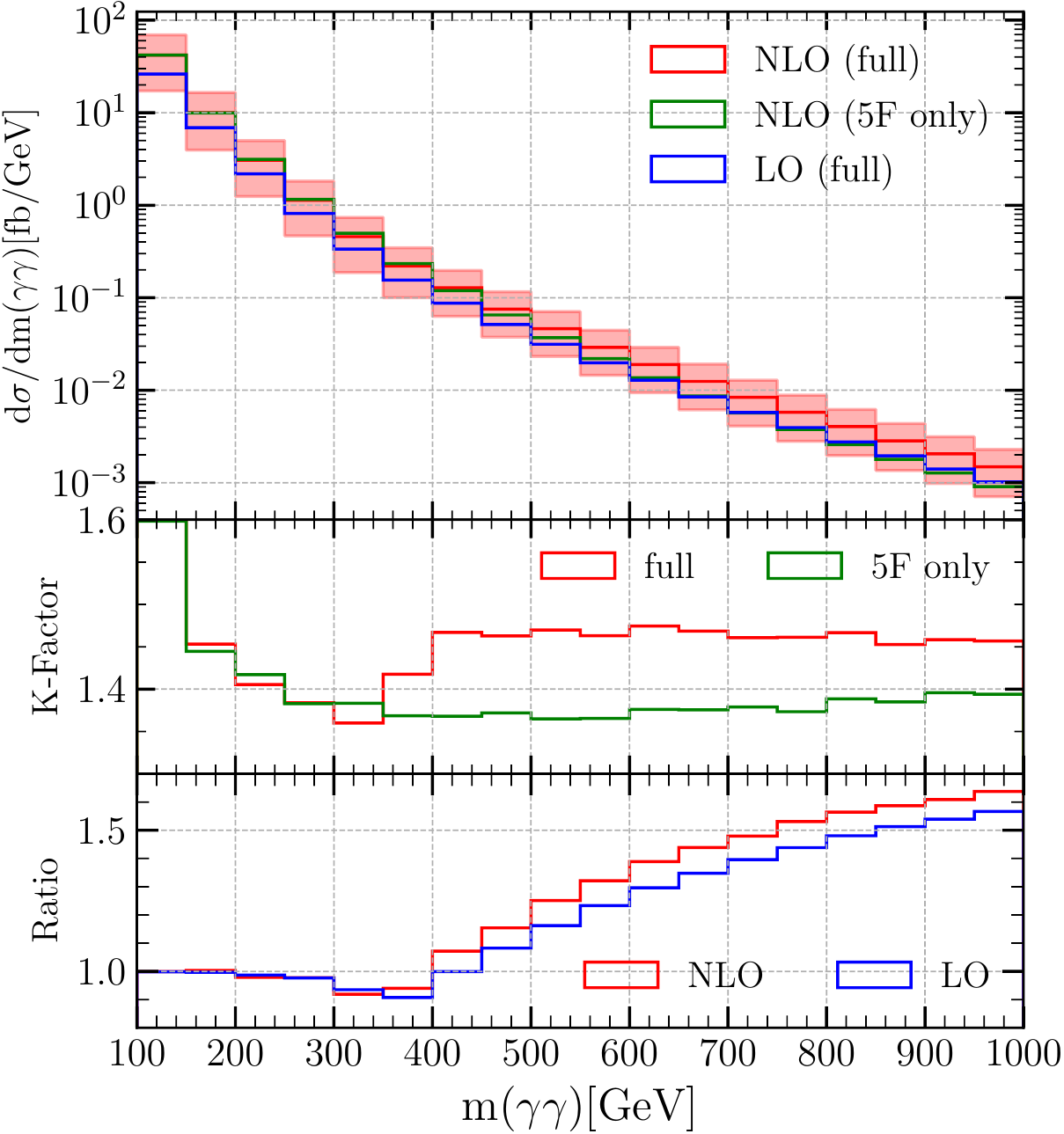}
	\caption{The differential cross section in $m(\gamma\gamma)$. 
	The cases of light quarks only and the full result (including top quark) are shown.
	The band indicates scale uncertainties for ``\text{NLO}~(full)''.
	The corresponding $K$-factor as well as the ratio between full and 5F are shown.}
	\label{fig:bin-cp}
\end{figure}

In the top inset of Fig.~\ref{fig:bin-cp}, we show the differential cross section at LO and NLO for the full result, {\it i.e.}, including five massless quarks along with the top quark (full) and the NLO result for the  massless quarks only  (5F only) averaged over 50 GeV wide bins. In the middle inset, we plot the $K$-factor~($K^{\text{NLO}}=\sigma^{\textrm{NLO}}/\sigma^{\text{LO}}$), which shows that NLO corrections for the top quark contribution in the high invariant mass regions are more important than those for the light quarks. Moreover, we plot the ratio between the ``full" and ``5F only" contributions at LO and NLO, in the lower inset. It clearly shows that the effect of the top quark mass is negligible in the low energy region both at LO and NLO as the top-quark contribution is parametrically suppressed as $\mathcal{O}(s^2/m_t^{4})$. As the energy increases towards the top-quark pair threshold,
the inclusion of the top-quark contribution leads to destructive interference, thus decreases the cross section. This behaviour is not affected by NLO corrections. However, as evident from the ratio as well as  the  $K$-factor plots, above threshold NLO corrections become large. Starting at about 400 GeV, the interference between light and top-quark contributions becomes constructive and since the NLO corrections for the top only case is larger than the light quark case, the full result displays a larger $K$-factor. As a consequence, at NLO the ratio between the full and 5F only results is larger, slowly approaching the predictions from the 6F (massless) calculation ($\approx1.86$).

In Table \ref{tab:dif-xs}, we provide benchmark values for the differential cross section.
As discussed before, in the low invariant mass region the top-quark contribution is tiny.
For example, at the Higgs mass region $m(\gamma\gamma)=m_H=125$GeV,
it is around $-0.1\%$. Going above top-quark pair threshold the top-quark contribution
decreases the cross section at LO, but the NLO cross section is almost unchanged.
Far above threshold, the interference turns to be constructive, and very slowly approaching the 6F (massless) limit.

{\bf Conclusions} -- In this letter, we have presented the first complete computation of the NLO corrections to $gg\to\gamma\gamma$ in the standard model, including both light-quarks and top-quark contributions. We have studied the top-quark effects in the total cross section and differentially, focusing on the invariant mass spectrum of the photons.  We find that the NLO corrections are important everywhere, but especially in the vicinity of  the top-quark pair threshold, where indeed an enhancement is expected on general grounds. A remarkable feature of the NLO spectrum is that the change of slope at the $t\bar t$ threshold becomes much more evident. Our calculation paves the way to improving the treatment of the threshold region at NLO including (pseudo-) bound state effects, with the goal to extract a short-distance top-quark mass, and to include background-signal interference effects at NLO accuracy in the production of new physics heavy scalar resonances decaying to diphoton final states.    

{\bf Acknowledgements} -- We thank Matteo Becchetti and Roberto Bonciani for providing us results of two-loop integrals for cross checks. We thank Claude Duhr and Pierpaolo Mastrolia for their advise. We are grateful to John Campbell for his help in comparing our code to the massless NLO computation in MCFM. This work has received funding by the Marie Sk\l{}odowska-Curie  MCnetITN3 (grant agreement no. 722104) and by the F.R.S.-FNRS with the EOS - be.h project n. 30820817. The work of M.K.M is funded by the UniPD STARS Grant 2017 ``Diagrammalgebra". Computational resources have been provided by the CISM at UCLouvain  and the C\'ECI funded by the F.R.S.-FNRS under convention 2.5020.11.

\bibliography{diphoton.bib}

\begin{thebibliography}{45}
\bibitem [{not()}]{note1}%
  \BibitemOpen
  \href@noop {} {\ }\bibinfo {note} {A comparison with our results has allowed
  to identify two bugs in the MCFM v8.0 implementation of minor numerical
  relevance, which have been corrected in MCFM v8.2. In addition the change of
  scheme to ``tH-V'' needs to be applied to obtain full agreement with our
  results. This will be included in the forthcoming release of
  MCFM.}\BibitemShut {Stop}%
\end{thebibliography}

\end{document}